\documentclass[sigconf]{acmart}

\AtBeginDocument{%
  }

\copyrightyear{2026}
\acmYear{2026}
\setcopyright{acmlicensed}
\acmConference[ITiCSE 2026]{Proceedings of the 31th ACM Conference on Innovation and Technology in Computer Science Education V. 1}{July 13--15, 2026}{Madrid, Spain}
\acmBooktitle{Proceedings of the 31th ACM Conference on Innovation and Technology in Computer Science Education V. 1 (ITiCSE 2026), July 13--15, 2026, Madrid, Spain}
\acmDOI{XXXXXXX.XXXXXXX}
\acmISBN{XXXXXXX.XXXXXXX}

\usepackage[UKenglish]{babel}
\usepackage{csquotes}
\usepackage{adjustbox}
\usepackage{xspace}
\usepackage{tabularx}
\usepackage{colortbl}
\usepackage{subfig}
\usepackage{multirow}
\usepackage{booktabs}
\usepackage{listings}
\usepackage{scratch3}
\usepackage{hyperref}
\usepackage[capitalise]{cleveref}
\usepackage{balance}
\usepackage{siunitx}

\setscratch{scale=.60}

\usepackage{pifont}
\newcommand{\cmark}{\ding{51}}  
\newcommand{\xmark}{\ding{55}}  

\newcommand\definetool[2]{\newcommand{#1}{{\textsc{#2}}\xspace}}
\newcommand\initialism[2]{\newcommand{#1}{{\textsc{#2}}\xspace}}
\definetool{\toolname}{MeowCrophone}
\definetool{\Scratch}{Scratch}
\definetool{\Blockly}{Blockly}
\definetool{\leila}{LeILa}
\definetool{\whisker}{Whisker}
\definetool{\litterbox}{LitterBox}
\definetool{\bastet}{Bastet}
\definetool{\scratchblocks}{scratchblocks}
\definetool{\gpt}{GPT-4.1}
\definetool{\openai}{OpenAI}
\initialism{\api}{api}
\initialism{\json}{json}
\initialism{\llm}{llm}

\colorlet{punct}{red!60!black}
\definecolor{background}{HTML}{EEEEEE}
\definecolor{delim}{RGB}{20,105,176}
\colorlet{numb}{magenta!60!black}

\lstdefinelanguage{json}{
    basicstyle=\normalfont\ttfamily,
    numbers=left,
    numberstyle=\scriptsize,
    stepnumber=1,
    numbersep=8pt,
    showstringspaces=false,
    breaklines=true,
    frame=lines,
    backgroundcolor=\color{background},
    literate=
     *{0}{{{\color{numb}0}}}{1}
      {1}{{{\color{numb}1}}}{1}
      {2}{{{\color{numb}2}}}{1}
      {3}{{{\color{numb}3}}}{1}
      {4}{{{\color{numb}4}}}{1}
      {5}{{{\color{numb}5}}}{1}
      {6}{{{\color{numb}6}}}{1}
      {7}{{{\color{numb}7}}}{1}
      {8}{{{\color{numb}8}}}{1}
      {9}{{{\color{numb}9}}}{1}
      {:}{{{\color{punct}{:}}}}{1}
      {,}{{{\color{punct}{,}}}}{1}
      {\{}{{{\color{delim}{\{}}}}{1}
      {\}}{{{\color{delim}{\}}}}}{1}
      {[}{{{\color{delim}{[}}}}{1}
      {]}{{{\color{delim}{]}}}}{1},
}

\usepackage{fbox}
\newcommand{\interpretation}[2]{
  \vspace{0.4em}
  \noindent
  \colorbox{gray!20}{%
    \parbox{.97\linewidth}{%
      \if #1\empty\else
      \textbf{#1.}
      \fi%
      #2
    }%
  }%
}%

\usepackage{xcolor}
\usepackage{colortbl}
\usepackage{wrapfig}

\newcommand*{\belowrulesepcolor}[1]{%
  \noalign{%
    \kern-\belowrulesep
    \begingroup
      \color{#1}%
      \hrule height\belowrulesep
    \endgroup
  }%
}
\newcommand*{\aboverulesepcolor}[1]{%
  \noalign{%
    \begingroup
      \color{#1}%
      \hrule height\aboverulesep
    \endgroup
    \kern-\aboverulesep
  }%
}

\begin{document}

\title{Voice-Controlled \Scratch for Children with (Motor) Disabilities}

\author{Elias Goller}
\email{elias.goller@uni-passau.de}
\affiliation{%
  \institution{University of Passau}
  \city{Passau}
  \country{Germany}
}

\author{Gordon Fraser}
\email{gordon.fraser@uni-passau.de}
\affiliation{%
  \institution{University of Passau}
  \city{Passau}
  \country{Germany}
}
\author{Isabella Graßl}
\email{isabella.grassl@tu-darmstadt.de}
\affiliation{%
  \institution{Technical University of Darmstadt}
  \city{Darmstadt}
  \country{Germany}
}

\begin{abstract}
  Block-based programming environments like \Scratch have become
  widely adopted in Computer Science Education, but the mouse-based
  drag-and-drop interface can challenge users with disabilities. While
  prior work has provided solutions supporting children with visual
  impairment, these solutions tend to focus on making content
  perceivable and do not address the physical interaction barriers
  faced by users with motor disabilities.
  To bridge this gap, we introduce \toolname, an approach that uses
  voice control to allow editing code in \Scratch. \toolname supports
  clicking elements, placing blocks, and navigating the workspace via
  a multi-modal voice user interface that uses numerical overlays and
  label reading to bypass physical input entirely.
  As imperfect speech recognition is common in classrooms and for
  children with dysarthria, \toolname employs a multi-stage matching
  pipeline using regular expressions, phonetic matching, and a custom
  grammar.
  Evaluation shows that while free speech recognition systems achieved
  a baseline success rate of only 46.4\%, \toolname's pipeline
  improved results to 82.8\% overall, with simple commands reaching
  96.9\% accuracy. This demonstrates that robust voice control can
  make \Scratch accessible to users for whom visual aids are
  insufficient.
\end{abstract}



\maketitle

\section{Introduction}

Programming helps children build problem-solving skills and
perseverance~\cite{bouck2022providing}, yet many children with motor
disabilities cannot fully participate in typical introductory
courses. \Scratch, the most widely used visual programming tool for
children~\cite{sharfuddeen2020visual}, relies on mouse-based
drag-and-drop interactions, which is challenging for learners with
limited fine motor control~\cite{zubair2018evaluating}.
While accessibility has improved significantly for users with visual
impairments~\cite{pires2020exploring,wang2019evaluating,morrison2020torino},
students with motor disabilities have been largely overlooked. The
fundamental interaction model of \Scratch, in which children have to
use a mouse to drag and drop code blocks, requires fine motor skills
that present a significant barrier for children with conditions such
as Cerebral Palsy~\cite{okafor2022helping}. Unlike text-based programming, which can often be
adapted with standard assistive technologies, the specific physical
requirements of a block-based drag-and-drop interface render standard
\Scratch inaccessible for this user group.

Enabling users with motor disabilities to use \Scratch requires a
\emph{voice user interface} (\emph{VUI})~\cite{myers2019impact}, but
to the best of our knowledge, there is no actively developed and usable
tool available.
Accessibility tools such as Myna~\cite{wagner2012programming,wagner2015programming} translated
voice commands into \Scratch actions, but has not been maintained
since 2015~\cite{ProjectMyna_2015}, does not support \Scratch~3.0, and
is not practical for classroom use~\cite{myna-prototype,Wagner_Grey}.
Existing work for \Blockly~\cite{blockly} cannot access key parts of
\Scratch such as stage and costume
editors~\cite{okafor2022voice,okafor2022helpingphd}.

We therefore introduce \toolname, a voice-controlled interface that
runs in parallel to \Scratch, allowing users to operate it entirely
without physical input. Drawing inspiration from professional
voice-coding tools like Talon~\cite{talon-voice} and
Serenade~\cite{serenade}, \toolname acts as an assistive layer that
translates spoken commands into actions within \Scratch.  Users can
place blocks, manipulate sprites, navigate menus, and manage variables
entirely by voice. Since \toolname is a separate application, it is
easier to maintain and ensures upgrade compatibility.

Designing voice control for \Scratch brings practical challenges:
\toolname must support navigation, block creation and manipulation,
parameter setting, and object management, while also handling noisy
environments and inaccurate transcripts. We implemented a vertical
slice that demonstrates feasibility across these core
interactions. The tool includes confirmations for critical actions,
handles low-confidence commands, and offers shortcuts and contextual
commands to reduce frustration.

Even though baseline speech recognition accuracy was low (46.4\%
overall and 27.1\% for German), reflecting authentic classroom
conditions, our matching pipeline improved accuracy to 82.8\% overall.
Due to the explorative nature of this work, \toolname is evaluated
against a set of 28 accessibility criteria based on
WCAG~\cite{wcag21}, NAUR~\cite{naur}, and prior work~\cite{masina2020investigating} to validate its design
foundations and support future studies with the target group.


\section{Background and Related Work}

\subsection{\Scratch: Visual Programming}

\Scratch is the most widely adopted visual programming tool in
education, particularly for children aged eight to
sixteen~\cite{sharfuddeen2020visual,richcoding}. Learners create
programs by assembling colourful, interlocking blocks, a design that
eliminates syntax errors and engages beginners. \Scratch has been
shown to effectively support interests among diverse learners as well as the development of problem-solving skills, computational thinking, and creative
thinking.~\cite{maraza2021towards,ccakirouglu2025mathematical,grassl2022gender}.

From a technical perspective, \Scratch
3.0\footnote{\url{https://github.com/scratchfoundation}} is built on
the \Scratch Virtual Machine (VM) and \Scratch-Blocks (a fork of Google
\Blockly). The VM manages the state of the \emph{Stage} and sprites, and
maps blocks to executable functions. \Scratch-Blocks renders the
graphical interface, handling user interactions such as dragging,
snapping, and deleting blocks. While this architecture provides a
robust visual metaphor, it enforces a mouse-centric interaction model,
as the VM relies on UI events triggered by the drag-and-drop
interface.

\subsection{Accessibility Barriers and Existing Tools}

Since \Scratch and similar block-based programming environments rely
heavily on visual cues and precise fine-motor control, they present
significant barriers for students with
disabilities~\cite{stefik2024lv, okafor2022helping}. Children with
motor impairments, such as those with cerebral palsy, often struggle
with the sustained mouse control required to drag blocks across the
screen~\cite{okafor2022helping}. Similarly, learners with learning or emotional
disabilities as well as ADHD may find the unstructured, cluttered interface
overwhelming~\cite{galeos2020developing,zubair2018evaluating,grassl2024coding,oliveira2019scratch}.

Several voice-assisted tools have attempted to bridge this gap, though
none provide a comprehensive solution for modern \Scratch:
Okafor et al.~\cite{okafor2022helping,okafor2022helpingphd} developed
a voice-controlled system for \Blockly that maps spoken commands to
block actions. However, this tool is deeply integrated into the
\Blockly core and lacks support for \Scratch-specific features, such
as the \emph{Stage}, costume editor, and sprite management, rendering it
unsuitable for the full \Scratch curriculum.
Myna~\cite{wagner2015programming} is a standalone voice bridge for
\Scratch. While Myna successfully demonstrated that grid-based voice
control could mimic mouse actions, it was designed for \Scratch 1.4
and is incompatible with the \Scratch 3.0 web-based architecture,
relies on outdated dependencies, and is no longer actively maintained.

Consequently, there is currently no publicly available,
classroom-ready \emph{VUI} that supports \Scratch 3.0. A new solution
is required that operates in parallel with the modern \Scratch editor,
handling not just code generation but also complex UI interactions
such as sprite manipulation and stage management.

\subsection{Speech Recognition for Children}

Automatic speech recognition (\emph{ASR}) accuracy is measured by word
error rate (\emph{WER}):
$\mathit{Substitutions} + \mathit{Insertions} + \mathit{Deletions}) /
\mathit{Total Words} \times
100$.\footnote{\url{https://www.assemblyai.com/blog/how-accurate-speech-to-text}}
Accuracy depends on factors such as background noise, audio quality,
age, accent, speaking speed, and pronunciation. Streaming recognition
is usually less accurate than batch
processing.\footnote{\url{https://voicewriter.io/speech-recognition-leaderboard}}

Reliable voice control requires accurate \emph{ASR}, yet commercial
engines continue to underperform with children’s
speech~\cite{bradley2025voice}. Key challenges include higher acoustic
variability, shifting formant frequencies due to vocal tract
development, and irregular
pronunciation~\cite{shivakumar2020transfer,potamianos2004robust,lee1999acoustics,gerosa2006acoustic}. These
difficulties are compounded for users with motor disabilities;
dysarthria (slurred or slow speech) affects up to 78\% of children
with cerebral palsy, reducing \emph{ASR} accuracy to
as low as
50–60\%~\cite{mayoclinic_dysarthria_symptoms_causes,ballati2018hey,mahr2020longitudinal,mei2020speech}.

To mitigate transcription errors, raw \emph{ASR} output often requires
post-processing. Phonetic matching algorithms, such as Double
Metaphone (for English) or Cologne Phonetics (for German), encode
words into phonetic keys to identify intended commands despite
mispronunciations~\cite{double_metaphone,postel1969kolner}. Fuzzy
matching utilises edit distance metrics (e.g., Levenshtein distance)
to quantify string similarity, allowing recovering commands from
imperfect transcripts~\cite{baeza1998fast,chaudhuri2003robust}.

\section{Approach}

\toolname is a \emph{VUI} designed to run in parallel with the
\Scratch editor, enabling users with motor impairments to control the
interface using spoken commands rather than mouse and keyboard
interactions. It operates on a command-remainder architecture: the
``command'' captures the user's intent (e.g., ``place'', ``delete''),
while the ``remainder'' supplies necessary parameters (e.g., ``block
5'', ``turn left'').
To ensure robustness against imperfect speech recognition, \toolname
employs a multi-stage matching pipeline that validates transcripts
against a fixed vocabulary and block grammars. The architecture
bridges the user's voice input and the \Scratch internals by
interacting directly with the \Scratch VM and the \Blockly rendering
engine, ensuring that voice actions are reflected accurately in the
project state.

\begin{figure}[t]
    \centering
    \includegraphics[width=\linewidth]{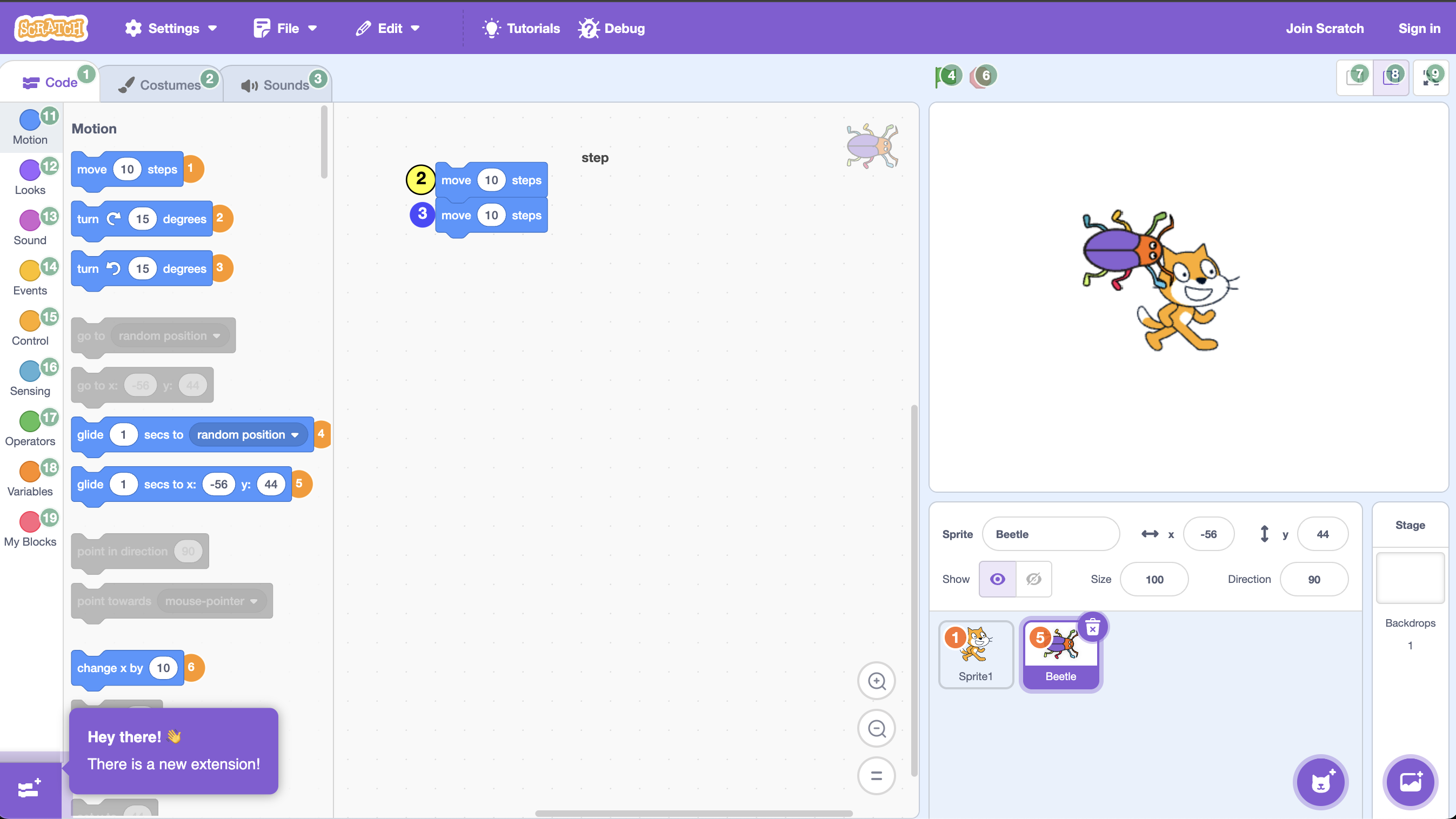}
    \caption{Numeric overlay for the workspace.}
    \label{fig:numeric_workspace}
\end{figure}
%

\begin{figure}[t]
    \centering
    \includegraphics[width=\linewidth]{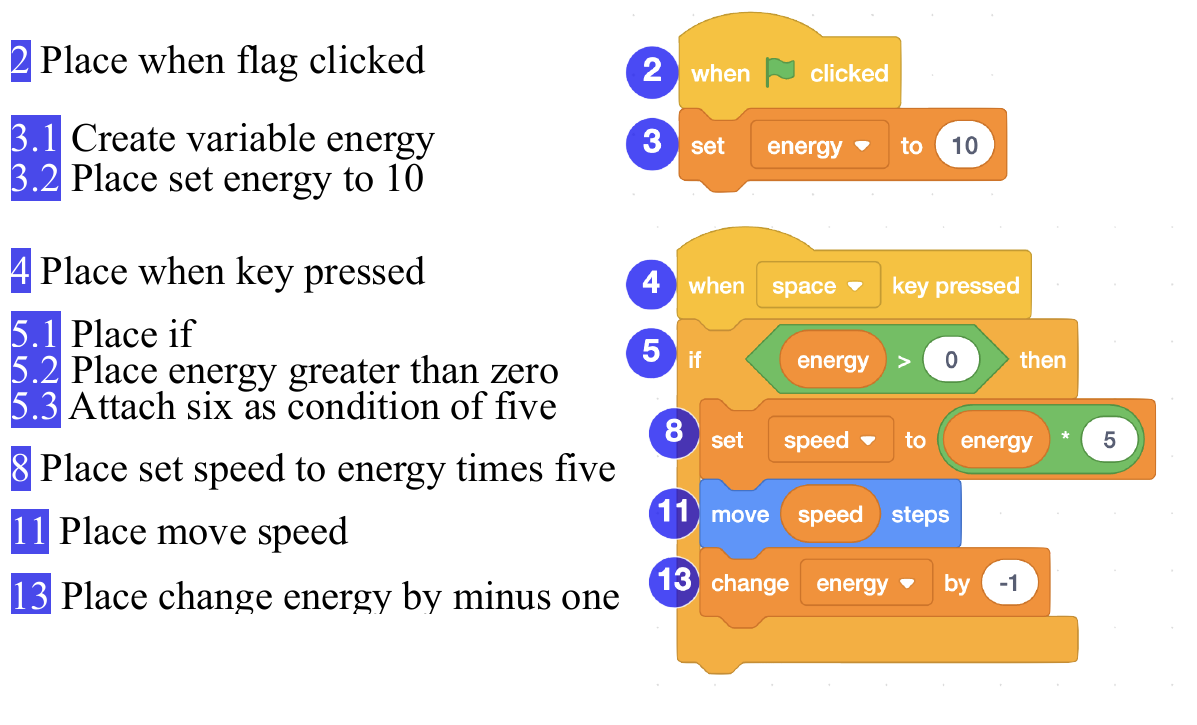}
    \caption{Voice commands mapped to \Scratch blocks.} 
    \label{fig:example}
\end{figure}

\subsection{Key Design Decisions}

The final system architecture evolved through several
iterations. Below, we summarise the central engineering challenges and
the design choices made to resolve them.

\paragraph{Transition from Simulated Mouse Control to Programmatic Block Manipulation}
Our initial prototype controlled \Scratch by simulating low-level
mouse clicks and key presses via a browser automation tool. While this
confirmed the viability of the interaction concept, it proved fragile;
UI automation broke easily depending on screen resolution, zoom
levels, and precise coordinate offsets required for stacking
blocks. Consequently, we replaced simulated inputs with programmatic
manipulation, utilising the internal APIs of \Scratch-Blocks
(\Blockly) to create, connect, and delete blocks directly. This
approach decoupled execution from visual layout, significantly
improving reliability.

\paragraph{Adoption of the \Scratch VM as Authoritative State}
While manipulating \Blockly provided robust placement, we encountered
synchronisation issues with shadow blocks. Shadow blocks are
non-draggable placeholders that provide default values for block
inputs (e.g., the number bubble inside a ``move steps'' block). We
discovered that when creating blocks programmatically via \Blockly,
the properties of these shadow blocks did not reliably propagate to
the \Scratch VM. As a result, shadow blocks would become detached or
reset after project reloads or sprite switches.
To resolve this, we instantiate new blocks by manipulating the
\Scratch VM directly, ensuring the authoritative state is correct,
while \Blockly is retained to handle interactions with existing
blocks.

\paragraph{Robust Matching Pipeline}
Early tests with the Web Speech
API\footnote{\url{https://developer.mozilla.org/en-US/docs/Web/API/Web_Speech_API}}
revealed that standard speech-to-text frequently returned inaccurate
transcripts. To mitigate this, we implemented a three-tier matching
pipeline. \toolname aggregates transcripts from interchangeable
Automatic Speech Recognition (ASR) services and processes them through
three stages:
\begin{enumerate}
\item Accurate Matching: Checks for exact command phrases.
\item Phonetic Matching: Uses algorithms like Double Metaphone to
  match similarly sounding words (e.g., recovering ``place'' from a
  transcript of ``plays'').
\item Fuzzy Matching: Uses edit distance as a final fallback for approximate matches.
\end{enumerate}
Hypotheses are scored and normalised into confidence values. This
allows \toolname to prioritise high-confidence commands while
gracefully recovering from ASR errors.

\paragraph{Grammar-Based Block Specification}
Users can place blocks by reading their text aloud (e.g., ``place turn
left by 10 degrees''). Initially, we used fuzzy matching for this, but
it proved unreliable for complex blocks with multiple inputs. To
address scalability, we introduced a grammar-based definition for all
supported blocks. Each block defines a grammar that expands into
regular expressions, complete with metadata for input types (e.g.,
variables, reporters). This trade-off adds implementation complexity
but ensures lightweight and accurate definition of block syntax,
allowing \toolname to distinguish between similar blocks effectively.

\paragraph{Interaction Modalities}
To accommodate different user needs, we moved beyond
``always-listening'' modes which were prone to accidental triggers. We
introduced Push-to-Talk and Toggle-to-Talk modes to give users
explicit control over input timing. Furthermore, to support users who
find reading block text difficult, we use Numerical Overlays
(e.g.,~\cref{fig:numeric_workspace},
\cref{fig:example}). These assign unique numbers to on-screen
elements (blocks, sprites, UI buttons), allowing users to execute
commands by saying ``Click 1'' or ``Select 5''. 
We implemented three overlay modes: \emph{Combined} (the main mode), which supports both spoken commands such as “place move twenty steps” and numerical commands like “place 10”; \emph{Smart}, which supports only spoken commands and was the original design; and \emph{Numerical}, which supports only numbers and can improve recognition accuracy because commands are shorter.

\subsection{Conceptual Model and Architecture}

The architecture of \toolname consists of a command processing
pipeline structured into distinct Layers and Domains.

\paragraph{The Processing Pipeline}

The pipeline drives the state machine, transforming raw audio into
actions. The primary states are:
\begin{itemize}
\item \textbf{Listening and Transcribing}: Audio is captured and
  converted to text candidates.
\item \textbf{Matching}: The pipeline applies the three-tier matching
  strategy to extract a Command (intent) and a Remainder (parameters)
  from the transcript.
\item \textbf{Processing}: The matched command is routed to a specific
  handler.
\item \textbf{Execution and Update}: The handler validates inputs,
  executes the logic, and triggers updates to the \Scratch VM,
  \Blockly workspace, or DOM.
\item \textbf{Feedback}: Visual confirmation is provided to the user.
\end{itemize}

\paragraph{Layers and Domains}

Rather than a strictly horizontal technical layering, the system uses
a domain-driven structure:
\begin{itemize}
\item \textbf{Core Layer}: Manages the global state, the matching
  pipeline, and the feedback system.
\item \textbf{Voice Layer}: Abstracts various ASR services (e.g., Web
  Speech API,
  Vosk\footnote{\url{https://www.assemblyai.com/blog/top-open-source-stt-options-for-voice-applications}})
  into a unified interface.
\item \textbf{Overlay Layer}: Manages the injection of SVG overlays
  into the DOM to render numeric labels on blocks and sprites.
\item \textbf{Domain Layers}: These contain the specific logic for
  different \Scratch features. For example, the Block Domain handles
  block placement and connections, while the UI Domain handles
  scrolling and button clicks.
\end{itemize}



\section{Evaluation}

To evaluate \toolname, we focused on three key dimensions:
feasibility, accessibility, and transcription reliability:

\noindent\textbf{RQ1}: Can common \Scratch features be supported via a \emph{VUI}?

\noindent\textbf{RQ2}: Does the \emph{VUI} comply with key accessibility standards?

\noindent\textbf{RQ3}: How accurate are voice transcription services for the \emph{VUI}, and does our pipeline improve them?

\subsection{Procedure}

We evaluated RQ1 and RQ2 using structured criteria derived from prior
work and accessibility standards, and RQ3 using a controlled
experiment with statistical analysis of real voice transcripts.

\subsubsection{Feature Support and Accessibility (RQ1 \& RQ2)}

To assess feature support (RQ1), we systematically tested essential
\Scratch actions such as block manipulation, sprite handling, and
navigation, rating each as fully supported, partially supported, or
not supported.
For accessibility (RQ2), we developed a catalogue of 28 criteria based
on the Web Content Accessibility Guidelines (WCAG)~\cite{wcag21} and
the Natural Language Interface Accessibility User Requirements
(NAUR)~\cite{naur}. These criteria cover feedback, timing, command
discoverability, and error handling. We evaluated the system against
these metrics using both manual inspection and automated checks.

\subsubsection{Transcription Reliability (RQ3)}
RQ3 examines whether our matching pipeline improves recognition over
raw transcripts.

\paragraph{Data Collection}

We designed a dataset of 24 spoken commands varying in complexity
(length and structure). These were recorded using two microphones, two
transcription services (Vosk\footnote{https://alphacephei.com/vosk/}
and Web
Speech\footnote{https://developer.mozilla.org/en-US/docs/Web/API/Web\_Speech\_API}
in Chrome), and two languages (English and German), resulting in 192
trials. Recordings were made at natural volume from a distance of
 50cm to simulate typical classroom
conditions.

\paragraph{Evaluation Conditions} 
To measure the impact of our system, we compared the raw transcription
output against the processed output of our pipeline. We defined four
evaluation conditions:
%
  
 \textbf{Baseline-Top}: A manual evaluation of the top hypothesis
  returned by the transcription service. This establishes how a basic
  system would perform without our pipeline. We applied lenient manual
  verification (e.g., accepting ``5'' and ``five'' as equal) to ensure
  we tested recognition quality rather than strict string formatting.

\textbf{Baseline-Any}: A manual evaluation checking if any alternative hypotheses returned by the service contained the
  correct intent. This represents the theoretical maximum accuracy of
  the raw ASR.

 \textbf{Pipeline-Top}: The transcript is processed by our
   pipeline (including block grammar matching). We count 
  success if the system's highest-confidence candidate triggers 
  correct execution.

 \textbf{Pipeline-Any}: The transcript is processed by our
  pipeline. We count a success if any of the ranked candidates
  generated by the pipeline would trigger the correct execution.
%

We enforce hierarchy constraints for data consistency: if a command
succeeds at Baseline-Top, it must succeed at Baseline-Any. The same
applies to the Pipeline conditions.

\paragraph{Statistical Analysis}

We report success rates, absolute/relative gains, and Cohen’s $h$ to
measure effect size (interpreted as small (0.2--0.5), medium
(0.5--0.8) or large (above 0.8).

To determine statistical significance, we use McNemar’s test for
paired comparisons (counting cases where the pipeline succeeds but the
baseline fails, versus the inverse). We apply the Holm correction to
control for multiple comparisons, setting $\alpha<0.05$.

Finally, we employ a mixed-effects logistic regression model to
analyse the 192 trials. This model estimates how factors like
microphone quality, service choice, language, and command complexity
affect the odds of improvement. We report odds ratios (values $>1$
indicate improvement). We intentionally omit interaction terms (e.g.,
Language$\times$Service) to avoid unreliable estimates caused by small
subgroup sizes (12 observations per cell).

\subsection{Features (RQ1) and Accessibility (RQ2)}

Before conducting user studies with children, which presents
significant challenges~\cite{henriques2024ethical,vsumak2023identification}, we needed to validate \toolname against
evaluation criteria found in accessibility research, most of which were derived from WCAG and NAUR.
We structured these criteria
following WCAG's format, describing expected behaviour and defining
passing requirements. \Cref{tab:evaluation-category} presents our
evaluation structured into ten categories, marking each criterion as
\emph{passed, partial or failed}. Importantly, using the voice
interface does not prevent users from exploring \Scratch's regular
drag-and-drop controls or keyboard shortcuts, and vice versa. This
allows teachers to assist children while they are using voice
commands~\cite{cheung2025teachers}.
%


%
\begin{table}[t]
\centering
\tiny
\caption{Criteria of the evaluation categories (RQ1, RQ2).}
\label{tab:evaluation-category}
\begin{tabular}{p{0.6cm}p{3.5cm}p{2.5cm}p{0.6cm}}
\toprule
No. & Name & Source & Pass \\ 
\midrule
 \multicolumn{4}{l}{Multi-Modal Input and Command Definition}\\
1.1 & Seamless Multi-Modal Interactions & WCAG 2.1.1, NAUR 13c, 3 & \cmark \\
1.2 & Speech Recognition Adaptability & NAUR 12b, \cite{masina2020investigating} & \xmark \\
1.3 & Confidence Thresholds and Confirmation & NAUR 13a, 13b & \cmark \\
1.4 & Custom Vocabulary Support & NAUR 18c & \cmark \\
1.5 & Use labels for unclear button names & WCAG 1.1.1, 2.4.6, 2.5.3, 4.1.2 & (\cmark) \\
 \midrule
\multicolumn{4}{l}{System Feedback and Error Handling}\\
2.1 & Multi-Modal Feedback & NAUR 4 & \xmark \\
2.2 & Clear Error Communication & WCAG 3.3.1, 3.3.3 & (\cmark) \\
2.3 & System Feedback Clearly Recognisable & \cite{masina2020investigating} & \cmark \\
2.4 & Undo and Redo Support & WCAG 2.5.2 & \cmark \\
 \midrule
\multicolumn{4}{l}{Timing and Flow of Voice Interaction}\\
3.1 & Adjustable Timing & WCAG 2.2.1 & \cmark \\
3.2 & Pause and Resume Functionality & WCAG 2.2.2 & \cmark \\
 \midrule
\multicolumn{4}{l}{Command Design and Discoverability}\\
4.1 & Simple Command Structure & NAUR 18b, \cite{masina2020investigating} & (\cmark) \\
4.2 & Command Discovery and Help & NAUR 16a, 16b, 16c & (\cmark) \\
4.3 & Multiple Command Phrasings & NAUR 18a & \cmark \\
 \midrule
\multicolumn{4}{l}{Focus Management and Input Compatibility}\\
5.1 & Visual Focus Indicators & WCAG 2.4.7 &  (\cmark)\\
5.2 & Screen Reader Compatibility & NAUR 10c, WCAG & (\cmark) \\
5.3 & Keyboard Shortcut Compatibility & NAUR 3, WCAG 2.1.1 & \cmark \\
 \midrule
\multicolumn{4}{l}{Context Handling}\\
6.1 & Step-by-Step Guidance & NAUR & \cmark \\
6.2 & Contextual Actions & NAUR & \cmark \\
 \midrule
\multicolumn{4}{l}{Personalisation and Adaptive Learning}\\
7.1 & User Profile Settings & NAUR 18c, 15 &  (\cmark)\\
7.2 & Multi-User Support & NAUR 12b, 18c & \xmark \\
 \midrule
\multicolumn{4}{l}{Privacy and Data Processing}\\
8.1 & Local Processing & VOSK feasibility & \cmark \\
8.2 & Data Control & Project constraint & \cmark \\
 \midrule
\multicolumn{4}{l}{Accessibility and Cross-Platform Support}\\
9.1 & User Testing with Disabilities & NAUR & \xmark \\
9.2 & Cross-Platform Compatibility & NAUR & \cmark \\
 \midrule
\multicolumn{4}{l}{Feature Coverage and Capability Levels}\\
10.1 & Core Feature Completion & NAUR 5 & \cmark \\
10.2 & Advanced Feature Completion & NAUR 5 & (\cmark) \\
10.3 & Expert Feature Completion & NAUR 5 & \xmark \\
\bottomrule
\end{tabular}
\end{table}

\textbf{Multi-modal input and command definition.}  We successfully
implemented independent overlay updates for voice controls across
sprites, blocks, library, and fullscreen states (\emph{1.1},
\Cref{tab:evaluation-category}). We built a confidence and
confirmation system with adjustable thresholds and three tiers
(\emph{1.3}), and made command aliases, remainders, and block grammars
configurable through external files (\emph{1.4}). We partially
addressed textual labelling by adding labels to clickable elements in
smart and combined modes, though we encountered a known bug preventing
labels from appearing on small buttons (\emph{1.5}). Criterion
\emph{1.2} is outside our project scope.

\textbf{System feedback and error handling.}  The feedback system
clearly communicates errors, although \toolname does not yet have
voice output, audio cues, or fix suggestions (\emph{2.1, 2.2}). We use
color to highlight command names and numbers, and push-to-talk and
tap-to-talk modes display a large red \emph{Recording} indicator
(\emph{2.3}). We maintained support for undo and redo functionality
(\emph{2.4}).

\textbf{Timing and interaction flow.} 
We made timers adjustable for confirmations, feedback, and speech recognition silence detection. Our push-to-talk and tap-to-talk modes help reduce time pressure (\emph{3.1}), and we provide explicit start and stop commands that allow users to pause when needed (\emph{3.2}).

\textbf{Command design and discoverability.} 
We generally kept commands simple or broke them into smaller steps, though some convenience options do add complexity; we have rated this as partial pending user testing (\emph{4.1}). We implemented tutorial systems but have not yet added a help command or contextual assistance (\emph{4.2}). We support multiple synonyms, flexible word orders, and various parameter phrasings to make commands more natural (\emph{4.3}).

\textbf{Focus management.} 
Sprites and blocks are highlighted when selected, but we have not yet added highlighting for workspace and flyout focus (\emph{5.1}). We designed the system to avoid interfering with \Scratch's screen reader compatibility, though we have not tested this thoroughly (\emph{5.2}). We ensured our voice interface additions do not conflict with \Scratch's existing keyboard shortcuts (\emph{5.3}).

\textbf{Context handling.} 
Our tutorial system guides users step by step through tasks (\emph{6.1}). The system uses focus to determine scroll targets, selection to enable property setting, and library context to change how it interprets commands (\emph{6.2}).

\textbf{Settings adjustable for personalisation.}  Settings can be
adjusted, but currently need to be set in configuration files directly
(\emph{7.1}). We have not yet implemented profile management, which would
require manual configuration (\emph{7.2}).

\textbf{Privacy criteria.} 
We met both by implementing Vosk for local speech processing and supporting multiple services (\emph{8.1, 8.2}). 

\textbf{Accessibility and cross-platform support.} We have not conducted a user study yet. One didactics expert on physical and motor disabilities provided expert feedback suggesting we show tutorial end results first, break tutorials into smaller steps, add voice feedback, and include supplementary keyboard controls (\emph{9.1}). We have verified the system works across multiple browsers and desktop operating systems using \emph{Playwright} (\emph{9.2}).

\textbf{Feature coverage.}  We implemented all core features in
\Scratch except menubar elements, for which we created dedicated
commands instead (\emph{10.1}, \Cref{tab:evaluation-category}). We
have implemented most advanced features, though some remain limited by
technical constraints (\emph{10.2}). We have not yet implemented the
features we categorised as optional enhancements (\emph{10.3}).

\subsection{Transcription (RQ3)}
\begin{table}[t]
    \centering
    \tiny
    \caption{Comparisons of metrics \emph{language}, \emph{complexity}, and \emph{ASR service} (RQ3). Gain is shown in points.}
    \label{tab:rq3-evaluation}
    \begin{tabular}{llrrrrr}
        \toprule
        Metric & Comparison & Base (\%) & Impr (\%) & Gain & Cohen's $h$ & McNemar $p$ \\
        \midrule
        \multicolumn{7}{l}{Overall Performance} \\
        & \emph{Base-Any vs Base-Top} & 46.4 & 68.2 & +21.9 & 0.45 (\emph{s}) & 4.5e-13 \\
        & \emph{Pipe-Top vs Base-Top} & 46.4 & 80.2 & +33.9 & 0.72 (\emph{m}) & 3.4e-15 \\
        & \emph{Pipe-Any vs Base-Any} & 68.2 & 84.9 & +16.7 & 0.40 (\emph{s}) & 4.4e-7 \\
        & \emph{Pipe-Any vs Base-Top} & 46.4 & 84.9 & +38.5 & 0.85 (\emph{l}) & 2.0e-20 \\
        \midrule
        \multicolumn{7}{l}{Language} \\
        EN & \emph{Base-Any vs Base-Top} & 65.6 & 77.1 & +11.5 & 0.25 (\emph{s}) & 9.8e-4 \\
        EN & \emph{Pipe-Top vs Base-Top} & 65.6 & 84.4 & +18.8 & 0.44 (\emph{s}) & 5.3e-4 \\
        EN & \emph{Pipe-Any vs Base-Any} & 77.1 & 87.5 & +10.4 & 0.28 (\emph{s}) & 3.1e-2 \\
        EN & \emph{Pipe-Any vs Base-Top} & 65.6 & 87.5 & +21.9 & 0.53 (\emph{m}) & 1.9e-5 \\
        \addlinespace
        DE & \emph{Base-Any vs Base-Top} & 27.1 & 59.4 & +32.3 & 0.66 (\emph{m}) & 1.9e-9 \\
        DE & \emph{Pipe-Top vs Base-Top} & 27.1 & 76.0 & +49.0 & 1.02 (\emph{l}) & 2.4e-12 \\
        DE & \emph{Pipe-Any vs Base-Any} & 59.4 & 82.3 & +22.9 & 0.51 (\emph{m}) & 6.0e-6 \\
        DE & \emph{Pipe-Any vs Base-Top} & 27.1 & 82.3 & +55.2 & 1.18 (\emph{l}) & 4.4e-16 \\
        \midrule
        \multicolumn{7}{l}{Complexity} \\
        Simple & \emph{Base-Any vs Base-Top} & 62.5 & 89.1 & +26.6 & 0.64 (\emph{m}) & 4.6e-5 \\
        Simple & \emph{Pipe-Top vs Base-Top} & 62.5 & 89.1 & +26.6 & 0.64 (\emph{m}) & 4.9e-4 \\
        Simple & \emph{Pipe-Any vs Base-Any} & 89.1 & 96.9 & +7.8 & 0.32 (\emph{s}) & 0.070 \\
        Simple & \emph{Pipe-Any vs Base-Top} & 62.5 & 96.9 & +34.4 & 0.96 (\emph{l}) & 9.5e-7 \\
        \addlinespace
        Medium & \emph{Base-Any vs Base-Top} & 42.2 & 60.9 & +18.8 & 0.38 (\emph{s}) & 4.9e-4 \\
        Medium & \emph{Pipe-Top vs Base-Top} & 42.2 & 84.4 & +42.2 & 0.91 (\emph{l}) & 3.4e-7 \\
        Medium & \emph{Pipe-Any vs Base-Any} & 60.9 & 89.1 & +28.1 & 0.68 (\emph{m}) & 3.6e-4 \\
        Medium & \emph{Pipe-Any vs Base-Top} & 42.2 & 89.1 & +46.9 & 1.05 (\emph{l}) & 5.6e-9 \\
        \addlinespace
        Complex & \emph{Base-Any vs Base-Top} & 34.4 & 54.7 & +20.3 & 0.41 (\emph{s}) & 4.9e-4 \\
        Complex & \emph{Pipe-Top vs Base-Top} & 34.4 & 67.2 & +32.8 & 0.67 (\emph{m}) & 3.9e-5 \\
        Complex& \emph{Pipe-Any vs Base-Any} & 54.7 & 68.8 & +14.1 & 0.29 (\emph{s}) & 0.070 \\
        Complex & \emph{Pipe-Any vs Base-Top} & 34.4 & 68.8 & +34.4 & 0.70 (\emph{m}) & 1.0e-5 \\
        \midrule
        \multicolumn{7}{l}{Service} \\
        Vosk & \emph{Base-Any vs Base-Top} & 45.8 & 62.5 & +16.7 & 0.34 (\emph{s}) & 3.1e-5 \\
        Vosk & \emph{Pipe-Top vs Base-Top} & 45.8 & 78.1 & +32.3 & 0.68 (\emph{m}) & 7.3e-8 \\
        Vosk & \emph{Pipe-Any vs Base-Any} & 62.5 & 82.3 & +19.8 & 0.45 (\emph{s}) & 3.1e-4 \\
        Vosk & \emph{Pipe-Any vs Base-Top} & 45.8 & 82.3 & +36.5 & 0.79 (\emph{m}) & 5.5e-10 \\
        \addlinespace
        Web & \emph{Base-Any vs Base-Top} & 46.9 & 74.0 & +27.1 & 0.56 (\emph{m}) & 6.0e-8 \\
        Web & \emph{Pipe-Top vs Base-Top} & 46.9 & 82.3 & +35.4 & 0.76 (\emph{m}) & 7.3e-8 \\
        Web & \emph{Pipe-Any vs Base-Any} & 74.0 & 87.5 & +13.5 & 0.35 (\emph{s}) & 2.3e-3 \\
        Web & \emph{Pipe-Any vs Base-Top} & 46.9 & 87.5 & +40.6 & 0.91 (\emph{l}) & 7.6e-11 \\
        \bottomrule
    \end{tabular}
\end{table}

To ensure data integrity, we first verified our dataset for hierarchy
constraint violations. Across all 192 trials, the ``Any'' metrics
consistently met or exceeded the ``Top'' metrics (e.g., Pipeline-Any
$\geq$ Pipeline-Top), confirming that our data is internally
consistent and correctly labelled.

\textbf{Overall Performance Improvement.}  The pipeline in \toolname
demonstrated a statistically significant improvement in command
execution success. By moving from a baseline that relies solely on the
top transcript to a pipeline that utilises any matched candidate,
success rates rose from 46.4\% to 84.9\%. \Cref{tab:rq3-evaluation}
details these comparisons across all categories.

\textbf{Impact of Language.}
While the pipeline benefited both tested languages, the magnitude of
improvement varied significantly.
English showed a solid improvement of 21.9 percentage points, rising
from a baseline of 65.6\% to 87.5\% (medium effect size, $h=0.53$).
German demonstrated the most dramatic gains. Starting from a low
baseline of 27.1\%, the pipeline boosted success to 82.3\% (large
effect size, $h=1.18$).
A mixed-effects model confirmed that language was the strongest
predictor of success, with German commands showing 4.7 times higher
odds of improvement (OR 4.72, 95\% CI 4.70--4.75, $p<0.001$). This
indicates the pipeline is particularly effective at compensating for
languages where the baseline ASR performance is lower.

\textbf{Effect of Command Complexity.} 
The pipeline proved effective across all levels of command complexity,
though the ``Medium'' category saw the greatest relative benefit:
Simple commands improved from 62.5\% to near-perfect execution at
96.9\% (large effect, $h=0.96$).  Medium commands achieved the largest
gain in the study, doubling success rates from 42.2\% to 84.4\% (large
effect, $h=1.05$).
Complex commands improved from 34.4\% to 68.8\%. While this remains
the most challenging category, the effect size was still medium
($h=0.70$).

\textbf{ASR Service Resilience.} 
A key finding is that the pipeline effectively equalises performance
between different ASR services. While the Web Speech API provided
better initial transcripts (higher baseline), our pipeline added
significantly more value to Vosk's output. Ultimately, both services
reached similar final performance levels: Vosk at 82.3\% and Web
Speech at 87.5\%.
This suggests that our fuzzy matching and phonetic algorithms are
particularly valuable when initial transcription quality is
lower. Statistical analysis using a mixed-effects model found no
significant service effect on the final outcome, confirming that
\toolname is robust regardless of the underlying speech engine.

\section{Discussion}

This study demonstrates the technical feasibility of a
voice-controlled interface for \Scratch, confirming that major
features can be effectively supported through a multi-modal approach
(RQ1). By implementing core functionalities, including block
manipulation, sprite management, and variable handling, we
met 15 of 28 standard accessibility
criteria fully (RQ2).

\paragraph{Technical Feasibility and Interaction Design}
The system’s success relied on integrating multiple interaction
mechanisms: simulated mouse controls, \Blockly integration, and direct
\Scratch VM interactions. A key finding was the effectiveness of the
multi-modal overlay system. By allowing users to reference visual
elements via voice while preserving standard mouse and keyboard
controls, the system enables seamless switching between input
methods. This is particularly valuable in classrooms,
allowing teachers to intervene manually without disrupting the child's
voice workflow~\cite{cheung2025teachers}.

\paragraph{Pipeline Performance and Reliability (RQ3)}
Our improved transcription pipeline significantly outperformed
expectations, proving that reliable execution is achievable even with
free, privacy-preserving services.
The pipeline raised overall execution success from a baseline of
46.4\% to 84.9\%. Improvements were most significant for German commands,
where the pipeline compensated effectively for lower baseline
transcription quality. 
The three-tier confidence system successfully balanced safety and
speed, requesting confirmation only for uncertain matches while
executing clear commands immediately.
Push-to-talk and tap-to-talk modes proved more reliable than
continuous listening, likely due to reduced background noise
interference

\paragraph{Practical Implications}
The observed gain of 38.5 percentage points demonstrates that
post-processing can compensate for poor ASR quality to a significant
degree when the domain grammar is known. Notably, simple commands
reached 96.9\% accuracy with the pipeline (up from a 62.5\%
baseline). While an 84.9\% overall success rate might seem
insufficient for frustration-free usage, this figure includes complex
commands. When resorting to the simpler numerical mode for block
placement rather than complex block grammar, most commands fall into
simple or medium complexity categories, achieving 96.9\% and 89.1\%
success respectively. Thus, practical success rates in a classroom
setting could be substantially higher. However, it must be noted that
children’s and dysarthric speech is challenging to transcribe,
potentially pushing error rates higher than those observed in our
controlled tests. Even at $\sim$89\% accuracy, one in ten commands
fails, which poses a risk of frustration.


\paragraph{Challenges and Technical Trade-offs}

Several approaches proved technically costly or less effective:
While enabling users to ``read code out loud'' (e.g., placing blocks
and parameters in one utterance) worked in testing, it added
substantial complexity and performance overhead. Weak transcripts
often contained unrecoverable errors; for example, ``place glide'' was
frequently transcribed as ``plays slide'' due to phonetic
dissimilarity.
Our initial reliance on Playwright for UI automation proved fragile,
forcing a mid-project pivot to direct \Scratch VM interaction.

\paragraph{Novelty}
\toolname distinguishes itself from voice-only systems by maintaining
a hybrid, multi-modal overlay system that works for non-English
languages. 
Most importantly, we demonstrated that voice control for complex
visual programming is feasible using free, locally-run services,
removing the barrier of expensive cloud subscriptions for schools.

\paragraph{Limitations}
The primary limitation of this work is the absence of user studies
with children. To avoid exposing the target demographic to early-stage
frustration~\cite{henriques2024ethical,vsumak2023identification}, we opted for a criteria-based evaluation to confirm
theoretical accessibility first. Consequently, our performance data is
derived from one speaker across 192 trials using two free services
(Vosk and Web Speech), without evaluating premium providers or actual
children's speech. Furthermore, operations requiring operating system
control, such as file uploads, remain outside the system's voice
control capabilities

\paragraph{Future Research}
The most critical next step is user testing with children,
particularly those with motor disabilities, to validate design
decisions and assess the impact on motivation and learning
outcomes. Future technical work should focus on (1) Feature Expansion,
such as adding support for costumes, sounds, and standard tutorials to
increase educational utility; (2) applying phonetic matching to
remainder extraction and tuning aliases for common errors (like the
``place glide'' issue); (4) investigating whether children prefer the
efficiency of numerical overlays over the \emph{smart mode} grammar, which
could allow for removing the complex grammar matching system entirely.

\section{Conclusions}

By introducing \toolname, this paper establishes that a
voice-controlled interface for \Scratch is not only technically
feasible but capable of supporting complex programming tasks through a
privacy-preserving, local architecture. By integrating a custom
post-processing pipeline with the Vosk engine, we demonstrated that
free, offline ASR can be enhanced to provide reliable command
execution, offering a viable alternative to costly cloud-based
services.
The resulting system is designed for the real-world classroom. It
complies with major accessibility standards (WCAG, NAUR) and adopts a
multi-modal approach where voice overlays coexist with traditional
inputs. This flexibility ensures that students with motor disabilities
can participate in creative coding while allowing for seamless teacher
assistance.
While technical limitations remain regarding specific OS-level
interactions, the foundation is laid. The critical next step is to
transition from system validation to user evaluation. Future studies
must assess the tool's intuitive value for children with motor
disabilities, validating its potential to make computing education
genuinely inclusive. 
\toolname is available at \url{https://github.com/se2p/MeowCrophone}.

\clearpage

\bibliographystyle{ACM-Reference-Format}
\bibliography{bib}

\end{document}